\documentclass{article}
\usepackage{spconf,amsmath,graphicx}
\usepackage{xcolor, url}
\usepackage{booktabs}
\usepackage{multirow}
\usepackage{tabularray}
\usepackage[subtle, mathdisplays=normal, floats=tight, tracking=tight, charwidths=normal, sections=normal]{savetrees}

\title{VCLIP: Face-based Speaker Generation by Face-Voice Association Learning}
\name{Yao Shi$^1$, Yunfei Xu$^1$, Hongbin Suo$^1$, Yulong Wan$^1$, Haifeng Liu$^2$}
\address{
    $^1$Data \& AI Engineering System, OPPO, Beijing, China\\
    $^2$University of Science and Technology of China, Hefei, China\\
    {\url{{shiyao1, xuyunfei, suohongbin, wanyulong, blade}@oppo.com}}
}
\begin{document}
\ninept

\newcommand{\draft}[1]{\textcolor{gray}{#1}}
\maketitle
\begin{abstract}
This paper discusses the task of face-based speech synthesis, a kind of personalized speech synthesis where the synthesized voices are constrained to perceptually match with a reference face image.
Due to the lack of TTS-quality audio-visual corpora, previous approaches suffer from either low synthesis quality or domain mismatch induced by a knowledge transfer scheme.
This paper proposes a new approach called Vclip that utilizes the facial-semantic knowledge of the CLIP encoder on noisy audio-visual data to learn the association between face and voice efficiently, 
achieving 89.63\% cross-modal verification AUC score on Voxceleb testset.
The proposed method then uses a retrieval-based strategy, combined with GMM-based speaker generation module for a downstream TTS system, to produce probable target speakers given reference images. 
Experimental results demonstrate that the proposed Vclip system in conjunction with the retrieval step can bridge the gap between face and voice features for face-based speech synthesis.
And using the feedback information distilled from downstream TTS helps to synthesize voices that match closely with reference faces. Demos available at \url{sos1sos2sixteen.github.io/vclip}.
\end{abstract}

\section{Introduction}
\label{sec:intro}

Current research in speech synthesis persues not only accurate and natural content delivery, 
but is also concerned with more advanced scenarios such as personalized speaker voice or style control.
Voice personalization was largely formulated as a voice cloning task in literature~\cite{tan2021survey}. 
but conditioning the synthesized voice on other modalities such as categorical labels~\cite{stanton2022speaker}, 
face imagery~\cite{goto2020face2speech,lee2023imaginary} or descriptive texts~\cite{yang2023instructtts} has attracted growing interest within the community, 
especially following the success of text-based image generation methods. 
Of particular interest is the face imagery modality among other cues for personalized synthesis voices, 
as speech and vision naturally co-occurs in our everyday experience and such paired data is easily obtained from large quantities of online videos~\cite{afouras2018lrs3,nagrani2017voxceleb,chung2018voxceleb2}. 
The connection between facial features and voice characteristics is two-fold.
Facial features reflect basic physiological conditions such as sex, weight and age, 
which contribute to the physical production of speech in the speakers' vocal organs~\cite{lammert2015short}.
Psychologically speaking, humans learn to associate certain face features with specific voice qualities through their experiences, which is known as face-voice concordance~\cite{smith2016concordant}.

A major discrepency between face and voice conditioning for speech synthesis is the quality of available datasets.
While synthesis task requires high quality audio for training, major audio-visual datasets such as LRS3~\cite{afouras2018lrs3} are generally noisy and cannot meet the requirements of mainstream TTS needs. 
Owning to these limitations, common approaches to face-based speech synthesis utilize a transfer learning scheme and attempt to learn mappings from face to voice features with neural networks. 
The transformed image features are used as proxies for speaker embeddings~\cite{goto2020face2speech,wang2022residual,lu2021face} during downstream TTS inference.  
However, synthesizing \textit{novel} speakers based on \textit{out-of-domain}~\cite{liang2022mind} features crosses two domain gaps.
And the learned feature mapping does not account for errors further accumulated after a face-based speaker embedding is predicted, causing a lack of coordination between the upstream mapping module and the downstream TTS module.
These design choices may contribute to performance degradation for TTS~\cite{wang23c_interspeech}.
Moreover, it is our opinion that predicting the \textit{true} speaker embedding by deterministic mapping of the reference image is essentially intractable, 
since face imagery alone does not provide enough information for reconstruction of a person's voice. 
Instead, different voices could be said to suit a single reference face equally well.
In this regard,  
VoiceMe~\cite{vanrijn22_interspeech} proposed using a human-in-the-loop sampling strategy to produce probable speaker embeddings for a face. 
This technique also takes the end-to-end synthesis result of the downstream TTS into account.
However, it scales poorly due to its requirement for intensive human interaction.

We therefore cast face-based speech synthesis as \textbf{a conditional speaker generation task~\cite{stanton2022speaker} aiming to produce probable candidate synthesis voices for reference faces}.
While \cite{vanrijn22_interspeech} leverages crowd-sourced human evaluators to link voices to faces, 
we aim to learn this association efficiently from openly available paired audio-visual data. 
We propose \textit{Vclip}, a CLIP-like contrastive learning~\cite{radford2021learning} scheme 
for distilling and connecting the semantic knowlege of a pretrained CLIP image encoder and a speaker encoder using low quality audio-visual data.
We then propose a retrival-based strategy for a GMM-based speaker generation model~\cite{turner2022generating} to produce probable speaker embeddings given reference images.
This procedure transfers the cross-modal association knowledge of Vclip into an existing zeroshot TTS system trained on high quality data. 
Furthermore, it can be extended to consider errors caused by the specific downstream TTS system in a straightforward fashion.
Experimental findings demonstrate that a retrival step is crucial in crossing the domain gap between face and voice features induced by a constrastive learning scheme~\cite{liang2022mind} in Vclip.
And feedback information distilled from the downstream TTS module helps to synthesize novel voices that matches more closely with reference faces.

The proposed Vclip system achieves 89.63\% AUC score on cross-modal verification for a Voxceleb testset~\cite{xiong2019voice}, surpassing a strong baseline method~\cite{chen2022self}. 
Using this system as guidance, we demonstrate the proposed strategy can produce synthesis voices that are highly correlated with reference faces, as evidenced by both perceptual and automatic evaluation studies.

\section{Background}
\label{sec:Background}

\textbf{Face-voice association.} 
Face and voice provide concordant and redundant information for a range of personal attributes such as 
levels of masculinity, age, weight etc. 
This establishes the real-world feasibility of static face-voice matching by human evaluators, as demonstrated in~\cite{smith2016concordant}. In the context of machine learning, Learnable PINs~\cite{nagrani2018learnable} proposed a cross-modal feature learning task to link face imagery to speech segments, known as face-voice association (FVA) learning.
In recent years, this task has drawn attentions from researchers~\cite{wen2021seeking,xiong2019voice}, with Self-Lifting~\cite{chen2022self} being among the state-of-the-art approaches. Self-Lifting employs an iterative learning process to capture the relationship between face and voice from unlabeled co-occuring audio-visual data, achieving SOTA results on Voxceleb~\cite{xiong2019voice}.

\textbf{Face-based speech synthesis.}
Unlike traditional FVA learning, which is formulated as a verification or retrieval task, Speech2face~\cite{oh2019speech2face} extends the problem to the generative domain, aiming to produce face images given speech features.
A major obstacle for reversing this process for TTS is the lack of high quality audio-visual data for training.
Face2Speech~\cite{goto2020face2speech} therefore couples a face-encoder's embedding space to an existing speaker encoder using SGE2E loss on noisy data. 
They suggest using the infered face feature in place of speaker embedding during synthesis,
effectively transfering the pairing knowledge learned from noisy data to TTS. 
Subsequent works~\cite{lu2021face, wang2022residual} follow this route by proposing different designs for the coupling module. 
In contrast, \cite{lee2023imaginary} trains TTS models with paired speech and face input on LRS3~\cite{afouras2018lrs3}, an audio-visual dataset for lip-reading. 
But the synthesis quality of this setup is severely limited by the poor quality of the corpus. 
Given no high-quality large-scale audio-visual corpus is openly available, we believe the transfer-based scheme is still a must for face-based TTS.

\textbf{Extending CLIP to the audio domain.} CLIP is an Internet-scale pretrained language-visual matching model with broad vision-related applications in both discriminative and geneartive fields~\cite{radford2021learning,gal2022stylegan}. 
Extending CLIP to other modalities such as audio has been proposed in literature.
AudioCLIP~\cite{guzhov2022audioclip} learns from co-occuring audio-visual data found in videos, producing a tri-modal embedding space useful for audio-visual event localization.
Wav2CLIP~\cite{wu2022wav2clip} uses a LiT-style scheme~\cite{zhai2022lit} to train an audio encoder against a CLIP image encoder for audio event classification. However, to the best of our knowledge, no prior work has linked CLIP features to speaker characteristics in speech.

\section{Method}
\begin{figure}[h]
    \centering
    \includegraphics[width=0.48\textwidth]{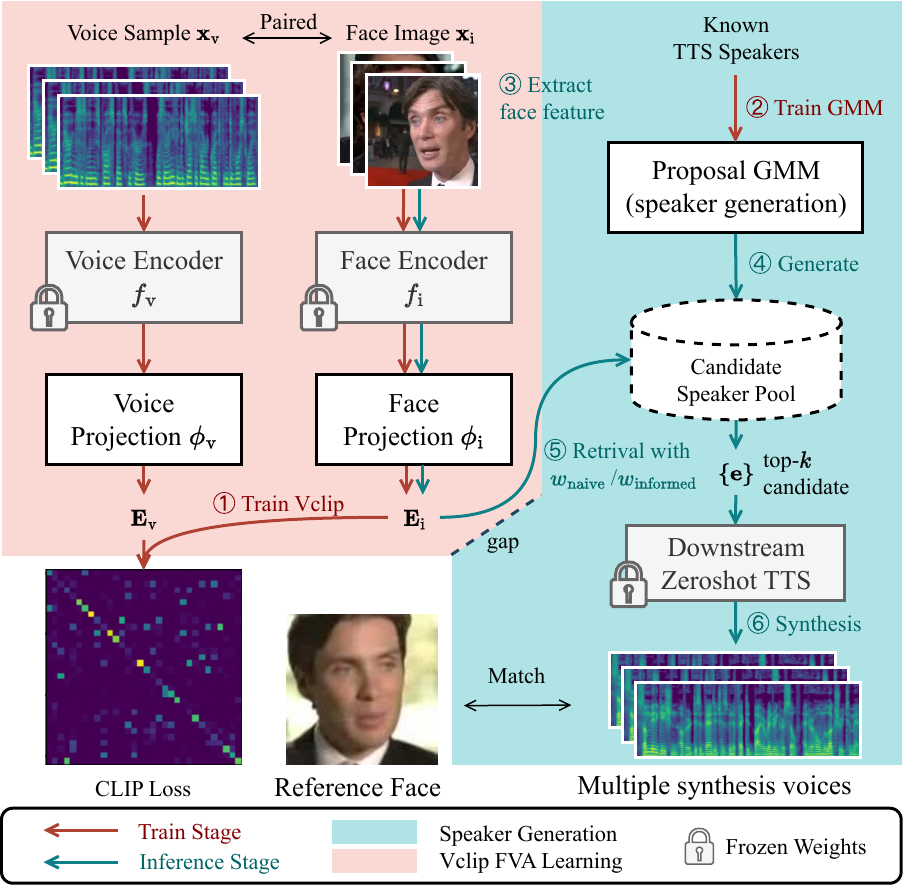}
    \caption{Overview of the proposed method. First a Vclip and speaker generation model is trained (red, steps 1-2). Then image features extracted from reference face image is used to retrive top-$k$ probable voices from a pool of generated candidate voices by using the trained Vclip as a zeroshot scoring function (blue, steps 3-6).}
    \label{fig:overview}
\end{figure}

We propose a two-part method for face-based speaker generation.
An overview of this method is illustrated in fig.\ref{fig:overview}.
In the first part, a matching system called Vclip is proposed to learn FVA knowledge from a large amount of audio-visual data found in online speech videos. 
In the second part, we propose a retrival procedure on novel speakers generated by a GMM-based speaker generation module. 
This method considers both the association knowledge provided by the trained Vclip module and the errors caused by the downstream TTS module to generate a set of high-quality target speaker embeddings for a reference face.

\subsection{Vclip for face-voice association}
Similar to Wav2CLIP~\cite{wu2022wav2clip}, Vclip is comprised of two separate modules for face (denoted $\text{i}$) and voice ($\text{v}$) modals. 
Each module contains a pretrained and frozen feature extractor network $f_\cdot(\cdot)$ for input, 
and a projection network $\phi_\cdot(\cdot)$ to further extract concordant information from respective features. 
We propose using a CLIP image encoder as our face encoder $f_\text{i}$, as CLIP is knwon to capture meaningful facial attributes~\cite{gal2022stylegan} and performs well on face recognition~\cite{bhat2023face}. 
A simple MLP is used as the face feature projection $\phi_\text{i}$. 
As for the voice encoder $f_\text{v}$, a pretrained speaker verification network is utilized.
We employ a flow module~\cite{DBLP:conf/iclr/DinhSB17} as the voice projection $\phi_\text{v}$ to transform
the distribution of pretrained speaker representations to that of the projected face features.

given a minibatch of $N$ pairs of face-voice samples $(\textbf{X}_\text{i}, \textbf{X}_\text{v})$, the extracted cross-modal representation of Vclip is given by $\mathbf{E}_\text{i}=\phi_\text{i}\circ f_\text{i}(\mathbf{X}_\text{i})$ and $\mathbf{E}_\text{v}=\phi_\text{v}\circ f_\text{v}(\mathbf{X}_\text{v})$ respectively. 
Here $\circ$ denote function composition.
We train Vclip using the CLIP constrastive loss given by eq.\ref{eq:clip-loss}, 
where $c = ||\mathbf{E}_\text{v}||_2||\mathbf{E}_\text{i}||_2$ represents the cosine similarity normalizer, $\text{tr}(\cdot)$ represents matrix trace, and $\sigma_\text{max}$ refers to the softmax function.
This formulation requires only co-occuring pairs, not speaker identity labels, which conforms to the \textit{unsupervised} setting in previous FVA learning works~\cite{chen2022self}.
\begin{equation}
    \label{eq:clip-loss}
    \textstyle
    \begin{aligned}
        \mathcal{L}&= \frac{1}{2N}(\text{tr}(\log\sigma_\text{max}(\frac{\mathbf{E}_\text{v}^T\mathbf{E}_\text{i}}{c})) + \text{tr}(\log\sigma_\text{max}(\frac{\mathbf{E}_\text{i}^T\mathbf{E}_\text{v}}{c})))
    \end{aligned}
\end{equation}
The CLIP loss encourages projected features to be similar in terms of cosine distance for positive pairs and vice versa. Therefore the learned projections are pushed to extract information that are concordant in both face and voice inputs.
But similarity does \textit{not} imply interchangablility.
It has been shown in~\cite{liang2022mind} that a modality gap exists for multi-modal models: 
Features from different source modals are always embedded in separate subspaces. 
Therefore we conjecture that treating face embeddings as a direct replacement for speaker embedding does not lead to optimal result for face-based TTS.

\subsection{Novel speaker generation with Vclip}
The trained Vclip model contains necessary information to link voices to faces.
But a modality gap means face embeddings extracted by Vclip is an out-of-distribution speaker embedding for TTS, which may degrade synthesis quality~\cite{wang23c_interspeech}. 
To address this problem, we propose a \textit{generate-and-retrive} strategy that produce samples with high correlation to the reference face inside the speaker embedding domain.

We start by modeling an unconditional distribution $p(\mathbf{e})$ for speaker embeddings of known speakers in the downstream zeroshot TTS embedding space using gaussian mixture models (GMM) following~\cite{turner2022generating}.
To obtain probable speaker embedding samples for a given reference face image $\mathbf{x}_\text{i}$, we employ Vclip as a zeroshot scoring function $w(\cdot; {\mathbf{x}_\text{i}})$ w.r.t. $\mathbf{x}_\text{i}$ for a pool of $N$ candidate samples $\{\mathbf{e}\}$ drawn from $p(\mathbf{e})$.
Samples with top-$k$ scores are selected as generated speakers.

\textbf{Naïve Scoring.} Assuming Vclip shares speaker representation with the downstream TTS model (use TTS speaker encoder as $f_\text{v}$), 
a naïve approach is to score the candidate samples $\{\mathbf{e}\}$ using the Vclip-projected $\phi_\text{v}(\mathbf{e})$ as the retrival index, as in eq.\ref{eq:naive-weight}.
\begin{equation}
    \label{eq:naive-weight}
    \textstyle
    w_\text{naive}(\mathbf{e};\mathbf{x}_\text{i}) := \cos(\phi_\text{v}(\mathbf{e}), \phi_\text{i}\circ f_\text{i}(\mathbf{x}_\text{i}))
\end{equation}

\textbf{TTS signature informed scoring. } It's worth noting that the naïve approach ignores the speaker mismatch between the input and output of zeroshot TTS systems~\cite{casanova2022yourtts}. 
That is, $f_\text{v}\circ\text{tts}(\mathbf{e})\neq\mathbf{e}$.
Here $\text{tts}(\cdot)$ denote the downstream TTS system whose input is the target speaker embedding, text input is ignored here for simplicity. In practice we may average on multiple sentences.
We may account for this deviation, and use brute-force durig scoring by considering $f_\text{v}\circ\text{tts}(\mathbf{e})$ as a replacement for $\mathbf{e}$ in eq.\ref{eq:naive-weight}. 
But performing synthesis on every candidate $\mathbf{e}$ during sampling incurs too much cost.
We note, given Vclip model, the composition $f_\text{v}\circ\text{tts}(\cdot)$ is determined solely by the TTS system.
This mapping is characteristic of the TTS system's input-output speaker mismatch phenomenon.
It captures inevitable voice cloning errors given input embeddings that is inherent to the TTS system in question.
Indeed, a perfect zeroshot voice clone would produce an identity mapping everywhere.
In light of this, we opt to distill an approximated version of this \textit{signature information} into a simple feedforward network $\text{s}(\cdot)$ with cosine embedding loss $\mathcal{L}_\text{signature} = \sum_{\mathbf{e}\in\mathcal{E}}\cos(\text{s}(\mathbf{e}), f_\text{v}\circ\text{tts}(\mathbf{e}))$.
We then define a TTS signature-informed scoring function $w_\text{informed}(\cdot)$ in eq.\ref{eq:s-informed-weight}.

\vspace{0em}
\begin{equation}
    \label{eq:s-informed-weight}
    w_\text{informed}(\mathbf{e};\mathbf{x}_\text{i}):=\cos(\phi_\text{v}\circ\text{s}(\mathbf{e}), \phi_\text{i}\circ f_\text{i}(\mathbf{x}_\text{i}))
\end{equation}



\subsection{Automatic evaluation of generated voices}
\label{sec:metrics}
We use a \textbf{separate} Vclip model as an automatic evaluator for assessing face-voice matching quality of generated voices to their reference faces. 
Given a pair $(\mathbf{x}_\text{i}, \mathbf{x}_\text{v})$ of face-voice sample, our method generates a set of $k$ probable speaker embeddings $\mathcal{E}$ conditioned on the face reference $\mathbf{x}_\text{i}$. 

\textbf{Voice reconstruction.} We define v2v (eq.\ref{eq:v2v}) as the cosine similarity between generated voices and the ground-truth voice of the reference face. 
This metric captures the ability to reconstruct the original voice behind a face from face image alone.

\begin{equation}
    \label{eq:v2v}
    \textstyle
    \text{v2v} := \frac{1}{k}\sum_{\mathbf{e}\in\mathcal{E}}\cos(
        f_\text{v}\circ\text{tts}(\mathbf{e}),
        f_\text{v}(\mathbf{x}_\text{v})
    )
\end{equation}

\textbf{Face-voice matching.} We define f2v (eq.\ref{eq:f2v}) as the cosine similarity between generated voices and their reference face. 
This metric reflects how well the proposed speakers in $\mathcal{E}$ matches with input face in terms of the learned association relationship of the evaluator.

\begin{equation}
    \label{eq:f2v}
    \textstyle
    \text{f2v} := \frac{1}{k}\sum_{\mathbf{e}\in\mathcal{E}}\cos(
        \phi_\text{v}\circ f_\text{v}\circ\text{tts}(\mathbf{e}),
        \phi_\text{i}\circ f_\text{i}(\mathbf{x}_\text{i})
    )
\end{equation}

\textbf{Speaker Generation Quality.} 
As statitics of generated speakers should be identical to real speakers for an ideal speaker generation system~\cite{stanton2022speaker}, 
we model the statistical characteristics of known TTS speakers' embeddings with a 4-component GMM.
The log-likelihood of speaker embeddings extracted from synthesized utterances w.r.t. this reference distribution serves as an indicator of the degree of domain-mismatch in TTS.

\section{Experiments}

\subsection{Experimental setup}
\textbf{Datasets.} For FVA learning, unless otherwise stated, we use videos from Voxceleb2-dev~\cite{chung2018voxceleb2} as training and development data for the proposed Vclip. 500 random speakers were kept from training and left for development.
For each video clip, we extract the first image frame and the entire audio track as paired face-voice data.
A preprocessed version of Voxceleb1~\cite{nagrani2017voxceleb} from \cite{chen2022self} is used for testing.
Note the exact trials used in \cite{chen2022self} are used for perforamance comparisons among FVA methods.
For zeroshot TTS, we use all 2130 train speakers from LibriTTS-R~\cite{koizumi2023libritts}, a high-quality restored version of LibriTTS, for training.

\textbf{Implementation.} For FVA learning, we use the ViT-B/32 CLIP released by openAI as the pretrained image encoder. And a H/ASP-based speaker verification model~\cite{heo2020clova} trained on Voxceleb2 is used as the pretrained voice encoder. Both encoders share a 512-dimensional embedding space. A ReLU-activated MLP~\cite{wu2022wav2clip} is used as our image projection layer ($\phi_\text{i}$) and a RealNVP~\cite{DBLP:conf/iclr/DinhSB17} flow is used as the voice projection layer ($\phi_\text{v}$).
All Vclip models were trained on a single NVIDIA V100 GPU with a batch-size of 1024, as we observe a performance increase when training on large mini-batches~\cite{zhai2022lit}.
The models are trained until AUC scores for the development-set starts to decrease.
For zeroshot TTS, we train a vanilla multi-speaker VITS~\cite{pmlr-v139-kim21f} for all our experiments in this chapter.
We follow \cite{turner2022generating} to implement a 100-component diagonal covariance GMM on principle components retaining the first 99\% data variance of knwon TTS speaker embeddings as our speaker generation system.

\subsection{FVA performance of Vclip}
We use Area Under ROC Curve (AUC) in cross-modal verification to assess the FVA performance of our Vclip model. Table~\ref{tab:auc} reports AUC scores for Vox1 testset. 
Only a train partition of Vox1 is used for the \textit{closed} setting. 
Under \textit{open} setting, Pins and Self-lifting is pretrained using AVSpeech~\cite{ephrat2018looking} and tuned on Vox1. 
As AVSpeech is not immediately available to us, Vclip uses Vox2 for pretraining, which is roughly at the same scale but has less identities.
The evaluation results demonstrate the proposed Vclip surpasses Self-Lifting and achieves best performance under both data settings. 
Consistent with \cite{bhat2023face}, our preliminary experiments show that the supervised face recognition models used in our baselines outperform CLIP in plain face recognition.
We attribute the performance gain observed here in FVA learning to the semanticly rich CLIP feature, suggesting it provides more concordant information with the voice features. 

We further investigate the impact of some design choices in Vclip. 
Table~\ref{tab:vclip-ablation} reports the AUC scores for the final and three variant versions of Vclip.
The most significant deviation occurs when the batch-size is reduced to 320, showing the CLIP loss requires large minibatches to gain perforamance~\cite{zhai2022lit}.
In constrast, replacing the CLIP loss with SGE2E loss~\cite{goto2020face2speech} with large batch-size also incurs a negative impact on the system.
This evidence supports the validity of the design choices we made for the proposed Vclip model.

\begin{table}
    \caption{Face-voice AUC results for Vox1-test}
    \label{tab:auc}
    \centering
    \begin{tabular}{lcc} 
    \toprule
    \multirow{2}{*}{system} & \multicolumn{2}{c}{AUC(\%)$\uparrow$}    \\
                            & \textit{closed} dataset & \textit{open} dataset  \\ 
    \midrule
    Learnable PINs~\cite{nagrani2018learnable}          & 81.88        & 84.70          \\
    Self-Lifting~\cite{chen2022self}                    & 86.70        & 89.40         \\
    \textbf{Vclip}                                     & \textbf{88.18}        & \textbf{89.63}       \\
    \bottomrule
    \end{tabular}
    \vspace{-1em}
\end{table}

\begin{table}[t]
    \caption{Ablation Study on Vclip.}
    \label{tab:vclip-ablation}
    \centering
    \begin{tabular}{lc} 
    \toprule
    system   & AUC(\%)$\uparrow$  \\ 
    \midrule
    \textbf{proposed} & \textbf{88.78}  \\
    $\text{bcsz}=320$    & 86.54  \\
    $-$CLIP loss    & 87.14  \\
    $-\phi_\text{v}(\cdot)$ flow    & 88.16  \\
    \bottomrule
    \end{tabular}
    \vspace{-1em}
\end{table}

\subsection{Automatic evaluation on face-based speaker generation}
\label{sec:facebasedtts-exp}
\begin{table}
    \caption[short]{Automatic evaluation results for generated voices.}
    \label{tab:auto-result}
    \centering
    \begin{tabular}{lccc}
        \toprule
        system &            v2v (±std) &        f2v (±std)$\uparrow$&         likelihood$\uparrow$\\
        \midrule
        ref value &         .588 (±.08)&         .307 (±.13)&                 557.40 (±22.42) \\
        \midrule
        w/o retrival &      .131 (±.09)&         .021 (±.21)&                 360.21 (±16.33) \\
        baseline &          .229 (±.09)&         .305 (±.13)&                 328.82 (±37.41)\\
        w/ $w_\text{naive}$& .249 (±.10)&        .301 (±.13)&                 \textbf{518.78 (±13.84)}\\
        \textbf{w/ $w_\text{informed}$}&    \textbf{.254 (±.10)}&     \textbf{.313 (±.13)}&                 516.22 (±15.31)\\
        \bottomrule
    \end{tabular} 
    \vspace{-1em}
\end{table}
We use the metrics described in section \ref{sec:metrics} to perform automatic evaluation on the generation quality of the proposed face-based TTS. A separate Vclip trained on Vox1 is used to calculate the scores.

\textbf{Baseline.} Since works on face-based TTS has been conducted under vastly different settings, 
fair comparisons cannot be performed for accurately reproduced implementations.
Following the main idea of~\cite{goto2020face2speech}, we implemented a \textit{supervised} variant of Vclip with SGE2E loss (Table~\ref{tab:vclip-ablation}, 3rd entry.) as a strong feature-mapping baseline. 
During generation, the projected face embedding $\phi_\text{i}\circ f_\text{i}(\mathbf{x}_\text{i})$ is directly used as the generated speaker embedding ($k=1$).
Compared to \cite{goto2020face2speech}, this baseline benefits from the use of CLIP features as well as a stronger zeroshot TTS system.

\textbf{Setup.} We sample $M=500$ positive face-voice pairs from Vox1 trials and evaluate averaged (over $k, M$) v2v, f2v and the generated speaker's log-likelihood for each system. We use a hparam of $k=10$ and $N=5000$ for retrival.
For ease of interpretation, we provide reference values derived from the same testset for each metric. 
The ref value for v2v is the cosine similarity of the TTS system on zeroshot voice cloning task.
And the ref value for f2v is the cosine similarity of ground-truth face-voice pairs. 
The ref likelihood values is the likelihood of known TTS speakers.

\textbf{Results.}
The results are reported in table~\ref{tab:auto-result}. 
We note all entries of v2v values are low compared to the true voice clone result on the same TTS system.
This indicates the generated voices are far from a prediction of the ground-truth voice behind a face in any case.
However, the f2v scores produced by our system are close, even surpass, their reference value, implying that meaningful \textit{matches} for the reference voice are generated.
A generate-and-retrive procedure with naïve scoring (w/ $w_\text{naive}$) performs on-par with our baseline in terms of f2v scores,
but has a close-to-reference likelihood value. 
This proves that a speaker generation based strategy for face-base TTS closes the domain gap induced by the mapping paradigm of baseline.
Moreover, by incooperating knowledge of the downstream TTS system with our proposed TTS signature informed scoring into the procedure (w/ $w_\text{informed}$), a higher f2v result is achieved.
This implies the feedback information distilled from the downstream TTS module complements our method and helps to retrive voices that matches more closely with reference faces.

\subsection{Subjective evaluation on face-based speaker generation}

We conduct subjective studies on the perceptual \textit{naturalness} and \textit{matchness} of our generated voices.
15 samples were evaluated by at least 12 evaluators for each entry in table~\ref{tab:mos-nat} and \ref{tab:mos-match}.

\textbf{Naturalness.}
We use a scale of 1-5 to measure the naturalness of our TTS under different settings.
We also include samples from Face-TTS~\cite{lee2023imaginary} directly trained on LRS3 for a quality comparison.
The evaluation results are reported in table~\ref{tab:mos-nat}.
We observe no significant deviation from plain voice-cloning occurs under our transfer learning setting. 
But Face-TTS performs significantly worse, suggesting face-based TTS still relies heavily on high-quality data for generating perceptually pleasing speech.

\textbf{Matching.}
The evaluators for FVA matching are tasked with scoring match-MOS (m-MOS), the degree of well-matching between pairs of face and voice stimuli.
We use a scale of 1-4, ranging from “not match” to “match well”, which is an inverted scale from the one used in \cite{goto2020face2speech}. 
The results are reported in table~\ref{tab:mos-match}.
To aid interpretation, multiple reference systems are constructed.
The audio stimuli for the \textit{ground truth} entries are selected from recordings in Vox1. 
The \textit{voice clone} contains pairs of face images and zeroshot speech cloned from true-match audios.
finally, \textit{ours} is a vclip w/ $w_\text{informed}$ system as described in section \ref{sec:facebasedtts-exp}.
We observe that m-MOS for $matched$ pairs are consistently above random-pairing baselines and the proposed method attains comparable results to the voice clone baselines with oracle information. 
This suggests that Vclip captures a true perceptually meaningful FVA knowledge between face image and generated speaker embeddings.

\begin{table}
    \caption[short]{Subjective naturalness results.}
    \label{tab:mos-nat}
    \centering
    \begin{tabular}{llc}
        \toprule
        task & setting & MOS (±$95\%$CI)$\uparrow$\\
        \midrule
        \multirow{2}{*}{voice clone} & known & 4.01 (±0.12) \\
        & zeroshot & 4.15 (±0.11) \\
        \midrule
        \multirow{2}{*}{face-based}& Face-TTS~\cite{lee2023imaginary} & 2.46 (±0.19) \\
        & ours & 4.14 (±0.10) \\
        \bottomrule
    \end{tabular}
    \vspace{-1em}
\end{table}

\begin{table}
    \caption[short]{Subjective face/voice matching results.}
    \label{tab:mos-match}
    \centering
    \begin{tabular}{lcc}
        \toprule
        \multirow{2}{*}{system} & \multicolumn{2}{c}{m-MOS (±$95\%$CI)$\uparrow$}\\
        & \textit{matched} pairs & \textit{random} pairs \\
        \midrule
        ground truth & 3.46 (±0.10) & 1.96 (±0.16) \\
        voice clone (oracle) & 3.33 (±0.10) & \multirow{2}{*}{2.31 (±0.17)} \\
        ours & 3.26 (±0.11) & \\
        \bottomrule
    \end{tabular}
    \vspace{-1em}
\end{table}

\vspace{-0.5em}
\section{Discussions}
\textbf{Relation to text modal.} As Vclip extends CLIP to the speech domain, 
it is conceptually viable to transfer prompts of physical descriptions to synthetic voice.
Preliminary experiments suggests that generated voices based on speaker prompts are sensitive only to simple concepts that reflect the target voice's gender. 
We find text conditioning without a training-time constraint lacking in controllability and requires further study.

\textbf{Attribute mismatch.}  
Although the proposed strategy effectively reduces the domain gap of generated speaker embeddings. 
An attribute-level mismatch is always present as audio-visual corpora collected in-the-wild contain variations in language, ethnicity and environmental factors that are lacking in reading-style TTS data.
As a result, although the proposed Vclip captures diverse variations across these factors (see our online demo). 
Only those attributes that coincides with LibriTTS are captured by an speaker embedding-centric TTS paradigm. 


\vspace{-0.5em}
\section{Conclusion}
We propose a two-part method for face-based TTS 
by formulating the problem as a conditional speaker generation task.
In the first part, a FVA method called Vclip is proposed.
Vclip utilizes the sematically-rich face representation of CLIP and a contrastive learning scheme to achieve 89.63\% AUC score on a public FVA learning benchmark.
In the second part, we transfer the FVA knowledge of Vclip to a GMM-based speaker generation model to produce highly probable novel speaker embeddings for an existing zeroshot TTS system. 
This conditional speaker generation procedure, combined with the feedback information distilled from the downstream TTS model, achieves superior matching results to a feature-mapping counterpart.
For future research, we plan to investigate more expressive downstream TTS methods to mitigate the attribute mismatch issue that we encountered with traditional multi-speaker TTS systems.


\bibliographystyle{IEEEbib}
\bibliography{strings,refs}

\end{document}